\documentclass[a4paper,11pt]{article}
\pdfoutput=1 
\usepackage{jinstpub}
\usepackage{siunitx}
\title{\boldmath Development of large area focal plane detectors for MAGIX}

\author[k]{P. G\"ulker} 
\author[k]{P. Achenbach}
\author[k]{S. Aulenbacher}
\author[k]{J. Bernauer}
\author[k]{S. Caiazza}
\author[k]{M. Christmann}
\author[k]{A. Denig}
\author[m]{S. Grieser}
\author[m]{A.-K. Hergem\"oller}
\author[m]{B. Hetz}
\author[m]{A. Khoukaz}
\author[k]{M. Klein}
\author[k]{T. Kolar}
\author[k]{M. Littich}
\author[k]{S. Lunkenheimer}
\author[k]{M. Mauch}
\author[k]{H. Merkel}
\author[k]{M. Mihovilovi\v{c}}
\author[k]{J. M\"uller}
\author[k]{J. Rausch}
\author[k]{Y. Schelhaas}
\author[k]{S. Schlimme}
\author[k]{S. \v{S}irca}

\author[p]{P. Bernhard}
\author[p]{A.S. Brogna}
\author[p]{Q. Weitzel}

\affiliation[k]{Institute for Nuclear Physics, Johannes Gutenberg University Mainz, Johann-Joachim-Becher Weg 45, 55128 Mainz, Germany}

\affiliation[m]{Institute for Nuclear Physics, University of M\"unster, Wilhelm-Klemm-Stra{\ss}e 9, 48149 M\"unster, Germany}

\affiliation[p]{PRISMA Detector Laboratory, Johannes Gutenberg University Mainz, Staudingerweg 9, 55128 Mainz, Germany}

\emailAdd{jguelker@uni-mainz.de}

\abstract{MAGIX is a planned experiment that will be implemented at the upcoming accelerator MESA in Mainz. Due to its location in the energy-recovering lane of the accelerator beam-currents up to \SI{1}{mA} with a maximum energy of \SI{105}{\mega\electronvolt} will be available for precision experiments. MAGIX itself consists of a jet-target and two magnetic spectrometers. Inside the spectrometers GEM-based detectors will be used in the focal plane for track reconstruction. The design goals for the detector modules are a spatial resolution of \SI{50}{\micro\meter}, a size of 1.20 $\times$ \SI{0.3}{\meter\squared} and a minimal material budget. To accomplish these goals we started developing several GEM-prototypes to study different behaviors and techniques to optimize the final detector design. The GEM foils used are provided by CERN and are trained, stretched and framed in our laboratory. The readout is done with an SRS based system. In this contribution the requirements, achievements and the ongoing developments are presented.}


 \collaboration{\includegraphics[height=17mm]{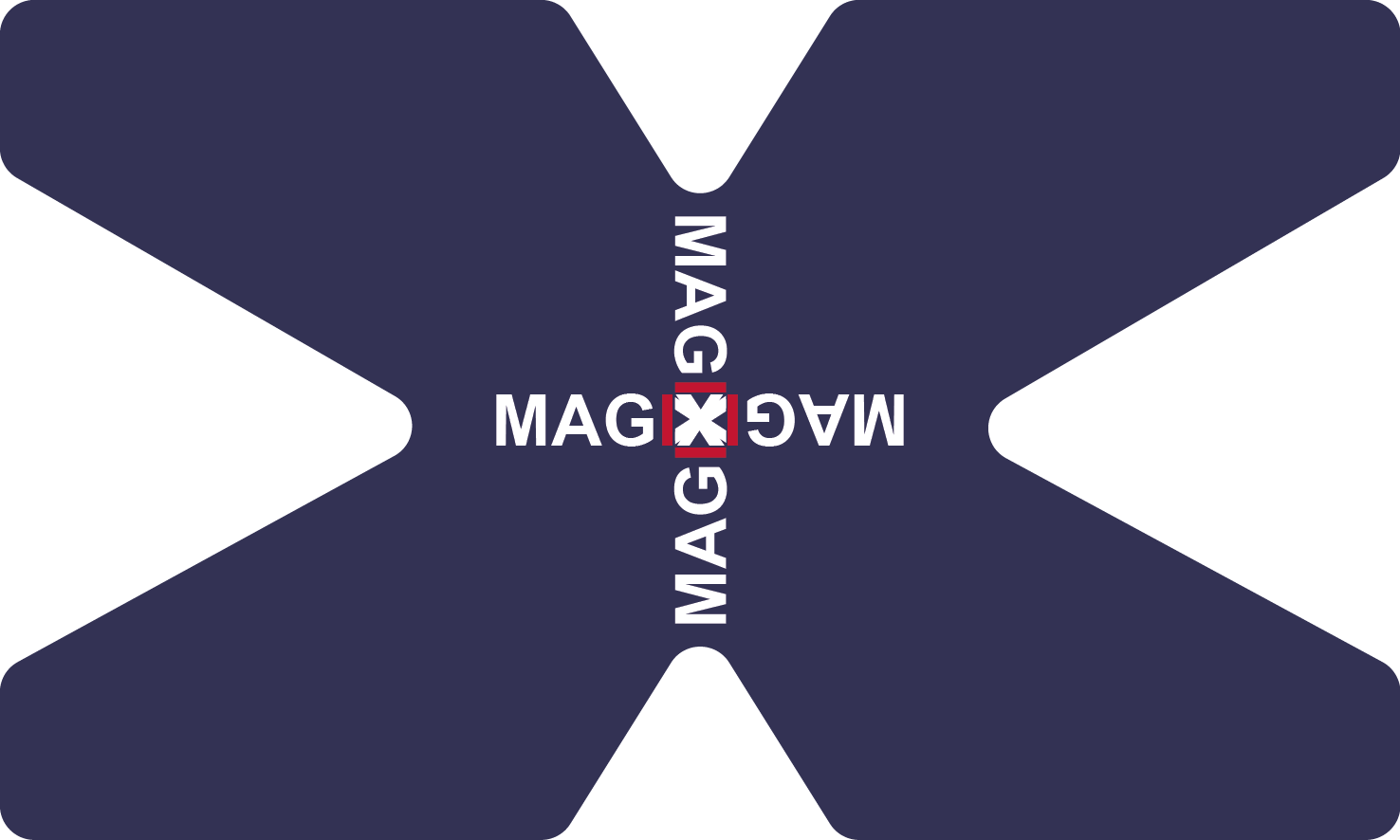}\\[6pt]
   MAGIX collaboration}
\proceeding{INSTR17\\
  February 27 - March 3. 2017\\
  BINP, Novosibirsk, Russia}

\begin{document}
\maketitle
\flushbottom

\section{Introduction: MESA/MAGIX} % (fold)

\begin{figure}[htbp]
\centering % \begin{center}/\end{center} takes some additional vertical space
\includegraphics[width=\textwidth,trim=30 110 30 30,clip]{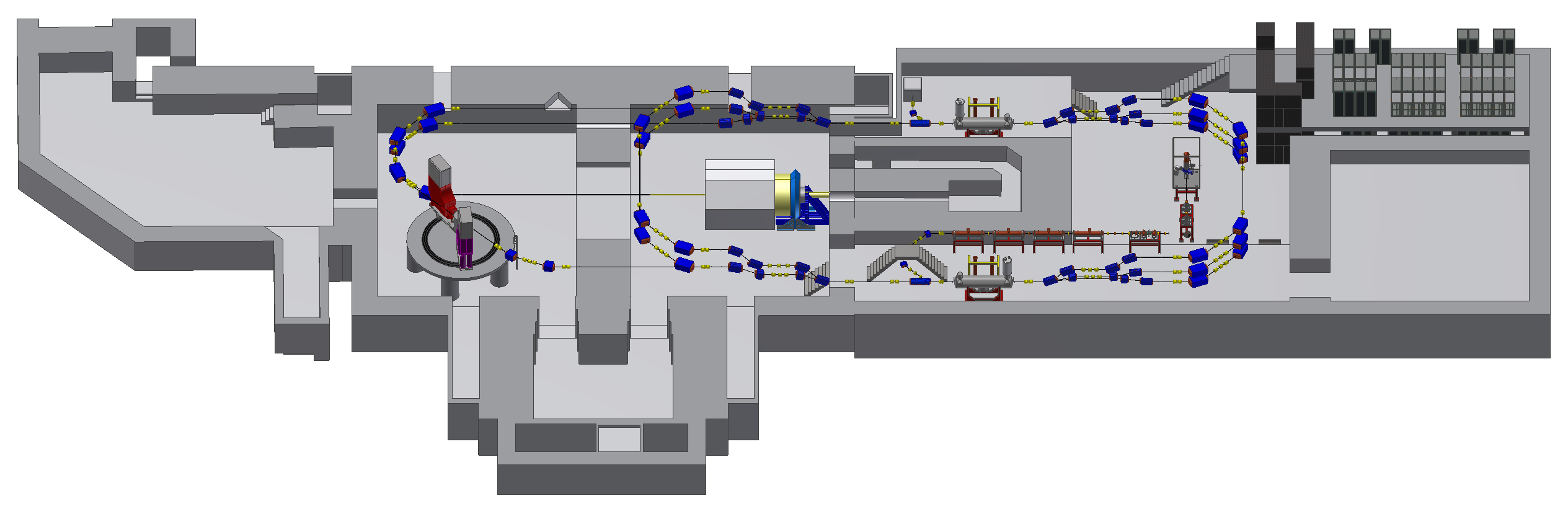}
\label{fig:mesa}
% "\includegraphics" from the "graphicx" permits to crop (trim+clip)
% and rotate (angle) and image (and much more)
\caption{Floor plan of the MESA-facility with the P2-Experiment shown in the middle and MAGIX on the left. The pre-accelerator, the superconducting cavities and the arcs may also be seen \cite{simon}.}
%TODO
%bessere name fuer arcs
%nummern ans bild? von wem ist das simon? wie heißt der noch gleich? daniel simon
\end{figure}

MESA\footnote{Mainz Energy Recovering Superconducting Accelerator} will be a new low energy electron accelerator built in Mainz, which unifies in a unique approach two different modes, the ERL\footnote{Energy Recovery Linac}- and the EB-mode\footnote{Extracted Beam} \cite{MESA13}. It will power two experiments, P2, as the user of the EB-mode, which will measure the weak mixing angle $\sin^2 \theta_W $, \cite{Berger:2015aaa} and the versatile MAGIX\footnote{MESA Gas Internal Target Experiment} experiment, which uses mostly the high intensity ERL-mode and will be explained here in more detail. In fig. \ref{fig:mesa} the floor plan of the accelerator and both experiments are shown.

MAGIX is perfectly suited for rare event searches as well as for nucleon and nuclei structure studies at the low-energy-frontier \cite{Merkel:2016bfj,Denig}. MESA will deliver an electron beam with \SI{20}-\SI{105}{MeV} and a current of \SI{1}{\milli\ampere} in the ERL-mode and possibly a polarized electron beam (with lower current) exploiting the EB-mode.
In order to perform precision measurements at this energy scale refined components, which meet challenging requirements are needed, e.g. a windowless target, two high resolution magnetic spectrometers and low material budget focal plane detectors. These parts will be described in the following sections.

\subsection{The Gas-Jet-Target} % (fold)
\label{sub:target}
One of the core-parts of MAGIX is the target. It is designed in such a way that neither the incoming particles, nor the scattered ones on their way to and through the spectrometers have to cross a window. This is accomplished by the use of a jet-target, \footnote{potentially even a cluster-jet-target} which utilizes a Laval nozzle to accelerate the gas and by that form a stable gas-jet that is injected into the vacuum inside the scattering chamber (see \cite{Taschner:2011ew}). An almost pointlike target with an integrated density of \SI{E19}{particles\per\centi\meter\squared} should be achievable, corresponding to a luminosity in the order of \SI{E35}{\per\second\per\centi\meter\squared} for a beam current of \SI{1}{\milli\ampere}. 

With this apparatus all gases from hydrogen up to xenon will be injectable, enabling to experiment on a lot of different nuclei.

\begin{figure}[htbp]
\centering \includegraphics[width=.4\textwidth,trim=30 30 30 30,clip]{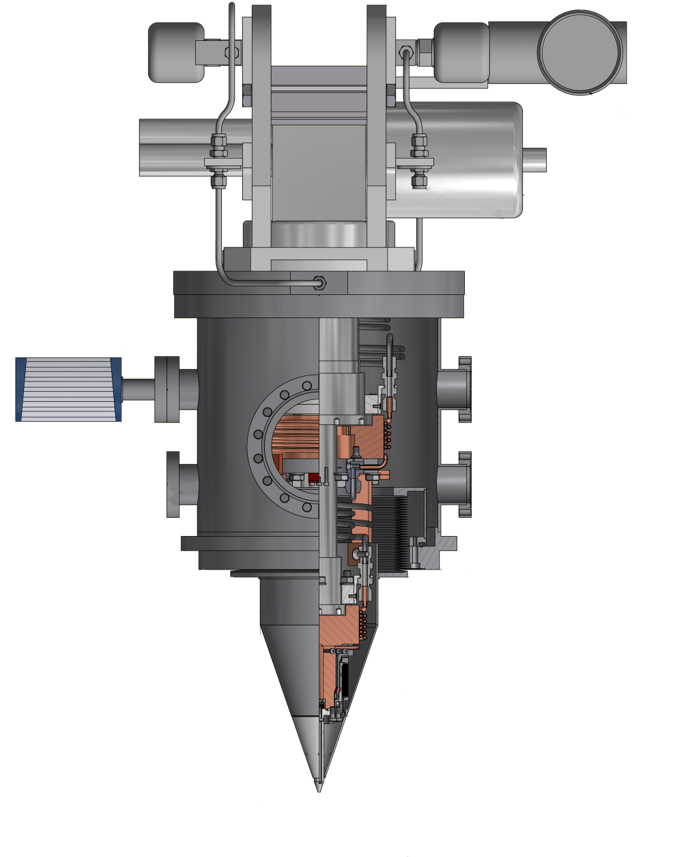}
\label{fig:jettarget}
\caption{Sketch of the MAGIX gas jet target setup. Mostly seen is the double stage cold head. On the bottom the Laval nozzle is shown \cite{jetTarget}.}
\end{figure}
% subsection target (end)

\subsection{Magnetic Spectrometers} % (fold)
\label{sub:spectrometers}

\begin{figure}[htbp]
\centering 
\includegraphics[width=.4\textwidth,trim=30 30 30 30,clip]{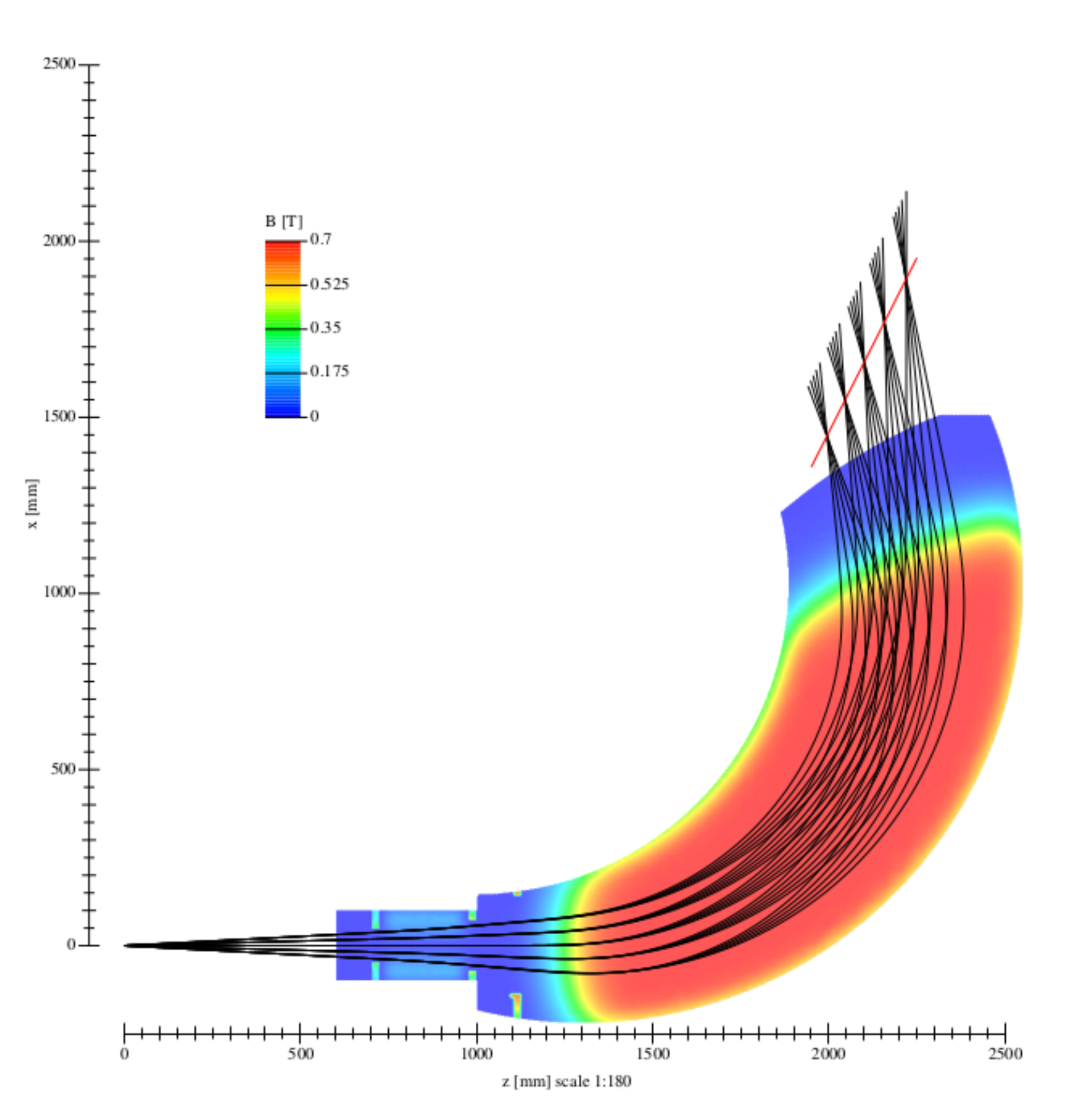}
\caption{Simulated field map of one spectrometer \cite{Mueller}. Particle trajectories for different momenta (in steps of \SI{5}{MeV}) and out-of-plane angles are shown, as well as the focal plane and the individual focusing points of particles with same momenta (regardless of the angle).}
\label{fig:fieldmap}
\end{figure}

To achieve a resolution in electron scattering cross sections of the order of 1\% an angular resolution for the electron detection of $\delta \theta < \SI{0.05}{\degree}$ is needed \cite{Merkel:2016bfj}. In addition an energy resolution of $\delta E < \SI{E-4}{}$ is required to resolve energy states when using nuclear targets in this low energy region \cite{Merkel:2016bfj}. This will be realized by two magnetic spectrometers, since the use of a toroid or solenoid was ruled out \cite{Molitor}.
A finite element simulation was already done for a quadrupole-dipole design with a requested straight focal plane and a detector system with \SI{50}{\micro\meter} spatial resolution  \cite{Mueller}. The resulting field map is shown in figure \ref{fig:fieldmap}, and the main parameters obtained from the simulation are given in table \ref{tab:spec}. The spectrometer design is laid out for a maximum central momentum of \SI{200}{MeV\per c} to be prepared for a potential energy upgrade of MESA.

The detector system will include scintillating detectors for trigger and timing purposes and GEM based focal plane detectors for tracking. As they will deliver tracks in the focal plane coordinate system, which have to be translated to the target system, a precise knowledge of the spectrometer optics is essential.

\begin{table}[htbp]
\centering
\caption{Parameters of the Spectrometers.}
\smallskip
\begin{tabular}{ll}
\hline
Height & \SI{1830}{\milli\meter}\\
\hline
Length & \SI{2800}{\milli\meter}\\
\hline
Max. central momentum & \SI{200}{\mega\electronvolt\per c}\\
\hline
Min. central angle & \SI{14}{\degree}\\
\hline
Momentum acceptance & 45 \%\\
\hline
Solid angle& 6.8 msr\\
\hline
Rel. momentum resolution &  $ <10^{-4}$\\
\hline
Angular resolution at target & $< \SI{0.9}{\milli\radian}$\\
\hline
\end{tabular}
\label{tab:spec} 
\end{table}

% subsection spectrometers (end)

% section introduction_mesa_magix (end)

\section{Focal Plane Detectors}
\label{sec:fp_detectors}
The design of the spectrometers requires to cover the whole focal plane ($\SI{1.2}{\meter}\times\SI{0.3}{\meter}$) with a detector system which allows to reconstruct particle tracks with a spatial resolution of \SI{50}{\micro\meter} to fulfill its design goal in momentum and angular resolution. Modern MPGDs\footnote{Micro Pattern Gaseous Detectors} may achieve this resolution \cite{Sauli:1999ya,Bressan:1998ji} if used in a well controlled setup.
Since MAGIX will be operated with currents up to \SI{1}{mA} particle fluxes of the order of \SI{}{MHz\per\centi\meter\squared} will be generated in the focal plane. In this harsh environment GEMs (\cite{Sauli:1996}) may be operated more stable compared to Micromegas and MWPCs \cite{Breskin:1974mm,Benlloch:1997we}.
In addition the low particle energy makes it necessary to minimize the radiation thickness as far as possible to avoid significant resolution reduction due to multiple scattering. This favors GEMs as well: Due to the construction principle it is easier to equip GEMs with a thin readout instead of Micromegas. It has to be explored if the detector should operate in a hodoscope- or a TPC-mode. We start following the more direct path of constructing a hodoscope as both detector designs require to master the GEM-foil preperation and handling anyway. 
 
The following section shows the construction of the first small prototypes to test the construction procedures, electronics and field configurations.

\section{Functional Tests}
The GEM foils, received from CERN, are trained (dried in nitrogen gas and if necessary high voltage cleaned) and the leakage current is measured in a first step. They are thermally stretched and framed in our laboratory. The sagging of the stretched foils is controlled and characterized with a profile-laser \cite{Prisma}. 
The framed foils are mounted on to a 2 $\times$ 256 crossed strips readout structure produced by CERN, or to a pad readout structure produced in our electronics lab. The detectors are powered with individual voltages by a CAEN module (A1580), which allows to monitor the current drawn with a resolution of \SI{50}{\pico\ampere}.
Connected to the strips are APV based frontend cards and an SRS data acquisition chain \cite{SRS}.
The trigger is mostly generated by two 10 $\times$ \SI{10}{\centi\meter\squared} scintillator paddles. Alternatively a pickup signal, taken from the last GEM-stage, is amplified and used for self-triggering purposes.

\begin{figure}[htbp]
\centering 
\includegraphics[width=.4\textwidth]{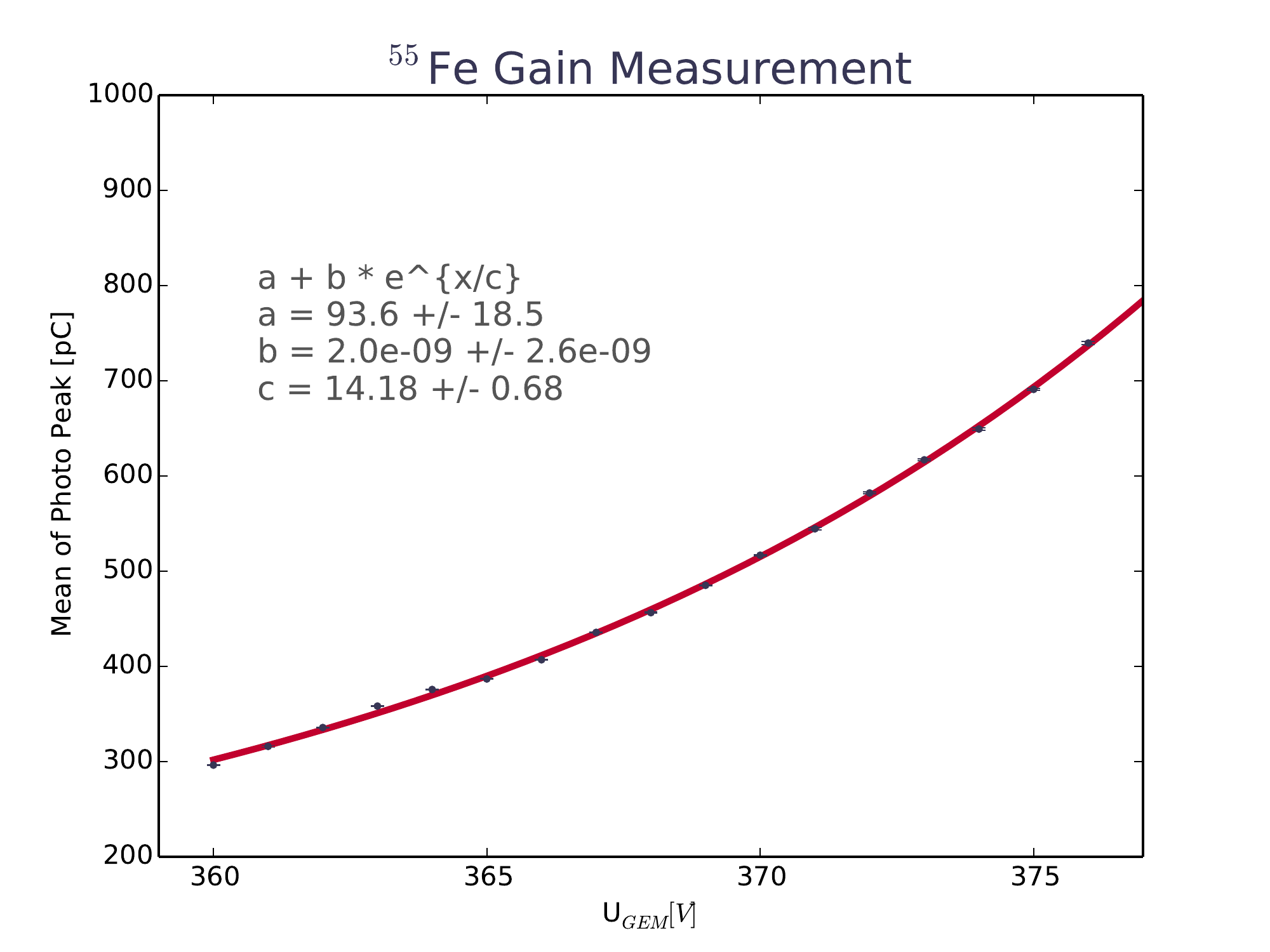}
\caption{Relative gain measured by varying all GEM voltages simultaneously and then plotting the charge of the photopeak. The used gas mixture is Ar:CO2 70:30.}
\label{fig:fe55}
\end{figure}

To test our first detector we measured the gain curve with a $^{55}$Fe-source, by taking the spectrum for different GEM voltages and plotting the charge of the photopeak (fig. \ref{fig:fe55}) versus the applied voltage. In the region between \SI{360}{\volt} to \SI{376}{\volt} the curve follows an exponential as expected \cite{SAULI20162}.

\begin{figure}[htbp]
\centering 
\includegraphics[width=.4\textwidth]{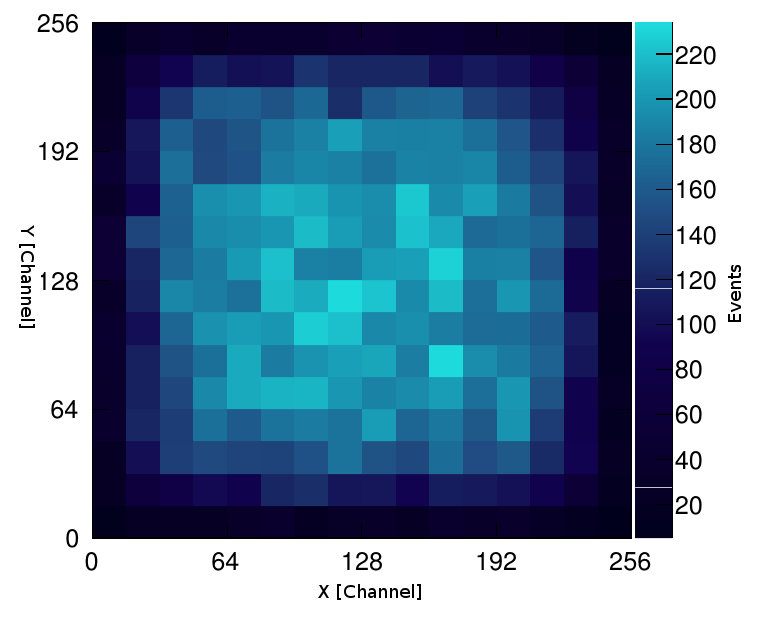}
\caption{Homogeneity measurement with cosmic rays.}
\label{fig:homo}
\end{figure}

The homogeneity of the detector area was measured with cosmic rays (fig. \ref{fig:homo}). The plot shows a smooth event collection over the whole detector, with a efficiency drop towards the edges. This is explained by geometrical effect of the scintillator paddles having the same surface as the active areas of the detector. The behavior at the edges will be studied in more detail in the future.

\section{Ongoing Activities \& Outlook} % (fold)
\label{sec:ongoing_activities}
Actually we are focusing our efforts on three different tasks:
The first one is to increase the detector size, in the next step to $\SI{30}{}\times\SI{30}{\centi\meter\squared}$. The foils are available and we will train them in the described way and try to stretch them thermally as well. The next step would be to either combine two of these modules, or to use even bigger GEM-foils and repeat the procedure. 

The second main task is that we have to decrease the radiation thickness by developing and equipping a detector with a thin foil based readout structure. For this we started implementing a one-layer structure consisting of strips in one direction and strips of pads (strads) in the other direction like done here \cite{Milner:2013daa}. 
In addition we try to make even the GEM foils thinner to reach the necessary resolution in the second layer of the hodoscope and not have it blurred by multiple scattering. For this we started characterizing foils with only several atomic layers of chromium as the conducting material on top and bottom of the GEMs. First of all the overall stability and resistance to sparks has to be checked.  

The third task is the improvement of the data acquisition and track reconstruction. As mentioned before our present setup is APV25-based, but we ordered several VMM3-chips, which will be tested extensively as soon as we get them. They should allow self-triggered data acquisition with rates up to \SI{1}{MHz}. We are also working on a track reconstruction software which is implemented in our MXWare framework.

When the new techniques are well under control we have to combine them and take the next step to our final size then.
% section ongoing_activities (end)

\section{Summary} % (fold)
\label{sec:summary}
The magnetic spectrometers of the upcoming versatile MAGIX experiment at MESA will be equipped with large area focal plane detectors exploiting the new possibilities of MPGDs. In Mainz we chose to develop GEM based detectors and were able to construct the first small prototypes and characterize them. With MAMI we have the great opportunity to perform regular electron-beam tests of the prototypes outside the lab. We already performed beam tests with out GEM-detectors, which look promising, but have to be fully analyzed.
The laboratory tests, two of them were presented here, worked out as expected. Now we are continuing our work to reach the final design goals of the large size, good resolution and small material budget.

% section summary (end)

% \bibliography{proceedingBINP}
% \bibliographystyle{naturemag}
% \bibliographystyle{plain}
% \bibliography{proceedingBINP}

% Please avoid comments such as "For a review'', "For some examples",
% "and references therein" or move them in the text. In general,
% please leave only references in the bibliography and move all
% accessory text in footnotes.

% Also, please have only one work for each \bibitem.

\newpage
% \listoftodos[Notes]
\end{document}